\documentclass[10pt,letterpaper]{article}
\usepackage{opex3}
\usepackage{graphicx}

\begin{document}

\title{\mbox{Revealing single emitter spectral dynamics} \mbox{from
intensity correlations} \mbox{in an ensemble fluorescence spectrum}}

\author{Xavier Brokmann$^{1}$, Lisa Marshall$^{2}$, Moungi Bawendi$^{2}$}

\address{$^{1}$20 rue Barban\`egre, F-75019 Paris. \\
$^{2}$Massachusetts Institute of Technology 77 Massachusetts Avenue,
Cambridge MA 02139. }

\email{xbrokmann@gmail.com} 



\begin{abstract*} We show that the single emitter linewidth underlying
a broadened ensemble emission spectrum can be extracted from
correlations among the stochastic intensity fluctuations in the
ensemble spectrum. Spectral correlations can be observed at high
temporal and spectral resolutions with a cross-correlated pair of
avalanche photodiodes placed at the outputs of a scanning Michelson
interferometer. As illustrated with simulations in conjunction with
Fluorescence Correlation Spectroscopy, our approach overcomes ensemble
and temporal inhomogeneous broadening to provide single emitter
linewidths, even for emitters under weak, continuous, broadband
excitation.
\end{abstract*}



\section{Introduction}
Single emitters often display dynamic and complex behaviors entirely
masked by ensemble measurements. Spectroscopy on these emitters has
reached impressive heights. There are systems, however, where it is
not feasible or desirable to separate the individual emitter from the
ensemble.  When confronted with the task of isolating individual
properties from large populations, spectroscopy offers a variety of
dedicated responses such as Doppler-free, hole-burning and photon-echo
spectroscopies \cite{SpecOverview} to resolve ensemble and temporal
averaging effects encountered in inhomogeneously and homogeneously
broadened samples. These diverse and powerful techniques share a
common trait; they all rely on the nonlinear optical properties of the
emitters. Hence, they tend to perform poorly on emitters with a small
or vanishing optical nonlinearity and questionably on samples too
delicate to handle the high excitation power required for nonlinear
optics.

Surprisingly, methods to extract the linewidth of a single emitter
from a broadened ensemble spectrum under the more gentle conditions of
linear excitation remain more elusive. In principle, the single
emitter linewidth can be determined from correlations among the
stochastic fluctuations in the ensemble emission spectrum
\cite{Koz01}. Progress in this direction was demonstrated in previous
works investigating the autocorrelation of the broad spectrum of
disordered nanostructures \cite{Sav01,Weg02}. Due to the long duration
necessary to record a high resolution spectrum, that approach cannot
resolve the fast temporal broadening effects found in most
inhomogeneous samples, severely limiting the ability to measure an
underlying single emitter linewidth.

In this Letter, we describe an experimental method revealing spectral
correlations of a single emitter with high spectral and temporal
resolution, despite the single emitter spectrum being obscured by an
ensemble emission spectrum. The approach - shown in Figure 1 and
reminiscent of our previous work on Photon Correlation Fourier
Spectroscopy \cite{Brok06} - consists of using a scanning Michelson
interferometer to turn spectral correlations in the broadened spectrum
into intensity correlations recorded by a Hanbury Brown Twiss
detection setup. Previously, we applied this method to an isolated
single emitter.  Here, we expand and generalize to obtain the same
dynamic single emitter spectral information from an ensemble of
emitters.  The method is investigated theoretically and demonstrated
with numerical simulations illustrating its significance in
conjunction with Fluorescence Correlation Spectroscopy (FCS).

\begin{figure}[t]
\centering\includegraphics{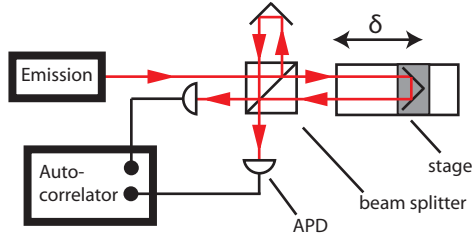}
\caption{Our approach uses an interferometer to convert fast spectral
fluctuations into intensity fluctuations. The cross-correlation of the
fluctuations in the interferometer outputs reveals single emitter
spectral dynamics with high temporal and spectral resolution.}
\end{figure}

\section{Intensity correlations in an inhomogeneous spectrum}

We introduce here the various quantities necessary to describe
intensity correlations in an inhomogeneous spectrum and then describe
a potential experimental setup dedicated to their measurement.

\begin{figure}[t]
\centering\includegraphics{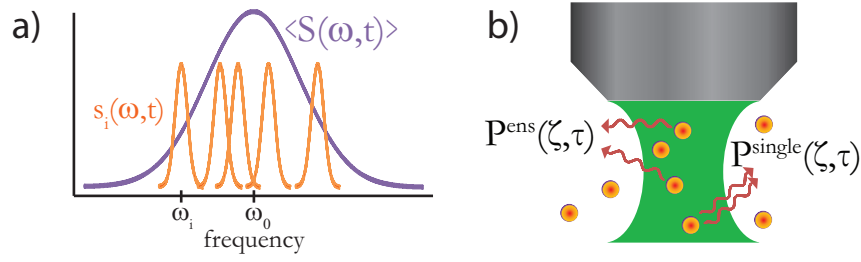}
\caption{Single-emitter and ensemble spectral correlations in the
fluorescence of a population of emitters. a) We first consider an
ensemble of emitters with the same narrow line shape but different
center frequencies. The ensemble spectrum would appear as a broad
Gaussian $\langle S(\omega,t)\rangle$.  b) With our setup, intensity
correlations from the emission of single emitters flowing under a
microscope objective will contribute to the single emitter spectral
correlation, $P^\mathrm{single}(\zeta,\tau)$, if the two photons
correlated are from the same emitter.  Correlations of photons
originating from different emitters will provide the ensemble spectral
correlation, $P^\mathrm{ens}(\zeta)$.}

\end{figure}

\subsection{Theoretical description}

We consider a collection of $N$ nearly identical emitters embedded in
a dynamic and spatially inhomogeneous environment. Due to slight
variations in the structure of the emitters (e.g. in their shape or
composition) and spatial inhomogeneity (as caused, for example, by
nanoscale disorder for emitters in a condensed medium), the spectra of
the single emitters do not collapse on a single narrow, homogeneous
stationary spectral line, but instead disperse their averaged
frequencies $\omega_1, ..., \omega_i, ..., \omega_N$ over a spectral
width $\Delta$ around some frequency $\omega_0$ (Figure 2a). Temporal
inhomogeneities additionally cause the lineshapes of the emitters,
$s_1(\omega,t), ..., s_i(\omega,t), ..., s_N(\omega,t)$, to be
explicitly dependent on time $t$, each emission line undergoing
independent, identically distributed temporal stochastic spectral
fluctuations over a mean-squared range $\sigma^2=\langle
\left[s_i(\omega,t)-\omega_i\right]^2\rangle$ centered around the
single emitter's average transition frequency $\omega_i$ (where
$\langle.\rangle$ denotes the average over many independent
observations).  At any time $t$, each emitter contributes to the
ensemble emission with an intensity $I_{i}(t)=\int
s_i(\omega,t)\mathrm{d}\omega$ and a normalized lineshape
$\hat{s}_i(\omega,t)= s_i(\omega,t)/I_{i}(t)$ (i.e.
$\int\hat{s}_i(\omega,t)\mathrm{d}\omega=1$).

Stochastic fluctuations in the total intensity of the spectrum
$I(t)=\sum_{i=0}^NI_i(t)$ are traditionally approached through the
second-order (intensity) correlation function $g^{(2)}(\tau)=\langle
I(t)I(t+\tau)\rangle/\langle I(t)\rangle^2$. Similarly, fluctuations
in the ensemble spectrum $S(\omega,t)=\sum_{i=0}^N s_i(\omega,t)$ can
be analyzed through the spectral correlation function $P(\zeta,\tau)$
:
\begin{equation}\label{FirstEq}
P(\zeta,\tau)=\langle\int
S(\omega,t)S(\omega+\zeta,t+\tau)\mathrm{d}\omega\rangle.
\end{equation}
Qualitatively, the spectral correlation function $P(\zeta,\tau)$
scales as the probability to measure a frequency difference $\zeta$
between two photons separated by a time interval $\tau$. Important
properties of $P(\zeta,\tau)$ appear when splitting up
Eq.\ref{FirstEq} into two distinct components as :
\begin{equation}
P(\zeta,\tau)=P^\mathrm{ens}(\zeta,\tau)+P^\mathrm{single}(\zeta,\tau)
\end{equation}
where $P^\mathrm{ens}(\zeta,\tau)=\sum_{i\neq j}\int\langle
s_i(\omega,t) s_j(\omega+\zeta,t+\tau)\rangle \mathrm{d}\omega$ and
$P^\mathrm{single}(\zeta,\tau)=N\langle\int
s_i(\omega,t)s_i(\omega+\zeta,t+\tau)\mathrm{d}\omega\rangle$. Since
emitters have statistically independent fluctuations, the component
$P^\mathrm{ens}(\zeta,\tau)$ rewrites as
$P^\mathrm{ens}(\zeta,\tau)=\sum_{i\neq j}\int\langle
s_i(\omega,t)\rangle\langle
s_j(\omega+\zeta,t)\rangle\mathrm{d}\omega$, i.e.
$P^\mathrm{ens}(\zeta,\tau)$ is independent of $\tau$ and reduces to
the autocorrelation of the average ensemble spectrum. Similarly,
$P^\mathrm{single}(\zeta,\tau)$ reduces to the autocorrelation of the
time-averaged single emitter spectrum as long as the single emitter
spectra $s_i(\omega,t)$ and $s_i(\omega,t+\tau)$ are uncorrelated,
i.e. for large values of $\tau$. But on shorter timescales $\tau\to
0$, temporal inhomogeneous broadening is suppressed as no spectral
fluctuations have time to occur, and $P^\mathrm{single}(\zeta,\tau=0)$
coincides with the autocorrelation of the single emitter spectrum. The
difference between $P^\mathrm{single}(\zeta,\tau=0)$ and
$P^\mathrm{ens}(\zeta)$ is illustrated in Figure 2b.

Information on the linewidth of the single emitter hence appears
systematically encoded into the spectral correlation function
$P(\zeta,\tau)$. As shown below, the spectral correlation function
$P(\zeta,\tau)$ turns out to be a quantity that can be measured
directly with a dedicated experiment from which the single emitter
linewidth can be extracted.

\subsection{Measurement setup}

The experiment involves a setup previously introduced to observe the
spectral dynamics of an isolated emitter at high temporal resolution
\cite{Brok06}. The emission is sent to a Michelson interferometer with
an arm continuously moving back and forth (at velocity $V$) over a
range of several fringes around an optical path difference $\delta$.
The intensities $I_\mathrm{a}(t), I_\mathrm{b}(t)$ at the outputs of
the interferometer oscillate as the Fourier transform of the emission
spectrum
\begin{equation}
I_{\mathrm{a,b}}(t)=\sum_{i=1}^N I_{i}(t)[1\pm\int_0^\infty
\hat{s}_i(\omega,t)  \cos(\omega\delta(t)/c)\mathrm{d}\omega],
\end{equation}
where $\delta(t)$ is the instantaneous optical path difference between
the arms ($\overline{\delta(t)}=\delta$). A pair of avalanche
photodiodes detects these intensities and a photon-counting board
computes their cross-correlation function $g^\times(\tau)$ :
\begin{equation}
g^\times(\tau)=\frac{\overline{I_a(t)I_b(t+\tau)}}{\overline{I_a(t)}\phantom{a}\overline{
I_b(t+\tau)}},
\end{equation}
where $\overline{\phantom{|}\ldots\phantom{|}}$ indicates
time-averaging over acquisition time of the intensity correlation.

Assuming the scanning speed $V$ is set low enough to ensure that
fringes oscillate on the photodiodes with a temporal periodicity
$c/2\omega_0V$ larger than the timescales $\tau$ under investigations,
the time-averaged intensity cross-correlation function
$g^\times(\tau)$ measured at the output of the scanning interferometer
decomposes as
\begin{equation}
g^\times(\tau) = g^\mathrm{ens}(\tau)+ g^\mathrm{single}(\tau), \label{Invert1}
\end{equation}
with :
\begin{eqnarray}
g^\mathrm{ens}(\tau) & = & \frac{N-1}{N}
\left(1-\frac{1}{2}\mathrm{FT}[p^\mathrm{ens}(\zeta)]_{\delta/c}\right)
\\ \label{Invert2}
g^\mathrm{single}(\tau) & = & \left(g^{(2)}(\tau)-1+\frac{1}{N}\right)
 \times \left(1-\frac{1}{2}\mathrm{FT}[p^\mathrm{single}(\zeta,\tau)]_{\delta/c}\right)
\label{Invert3}
\end{eqnarray}
where $p^\mathrm{ens}(\zeta)=P^\mathrm{ens}(\zeta)/\int
P^\mathrm{ens}(\zeta)\mathrm{d}\zeta$ and
$p^\mathrm{single}(\zeta,\tau)= P^\mathrm{single}(\zeta,\tau)/\int
P^\mathrm{single}(\zeta,\tau)\mathrm{d}\zeta$ denote the normalized
spectral correlation functions of the ensemble spectrum and
single-emitter spectrum respectively.

\begin{figure}[tb]
\centering\includegraphics{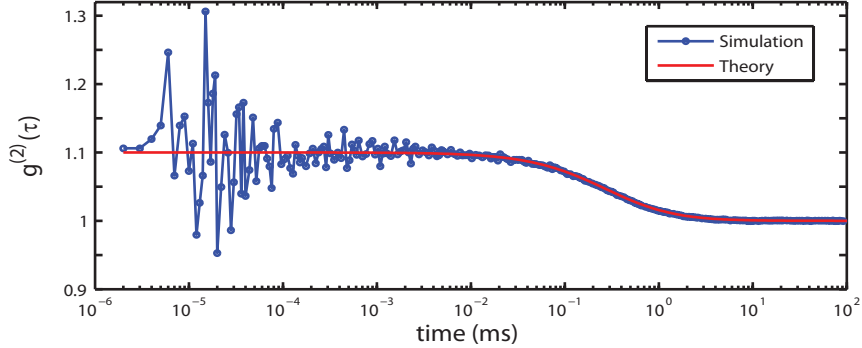}
\caption{The standard FCS intensity correlation function,
$g^{(2)}(\tau)$. Our method works in situations like FCS, where the
ensemble emission exhibits an intensity correlation function
$g^{(2)}(\tau)\neq1$ at short timescales. In these cases, the single
emitter spectral correlation, $p^\mathrm{single}(\zeta,\tau)$, is
weighted by $g^{(2)}(\tau)-1$ and can be separated from the background
ensemble spectral correlation, $p^\mathrm{ens}(\zeta)$.}
\end{figure}

This result simplifies in a few important cases. For an isolated
emitter ($N=1$), the cross-correlation function $g^\times(\tau)$
reduces to its single-emitter component, i.e.
$g^\times(\tau)=g^\mathrm{single}(\tau)$. Assuming the emission obeys
Poissonian statistics (i.e. $g^{(2)}(\tau)=1$), the intensity
cross-correlation function reads :
\begin{equation}
g^\times(\tau)= 1-\frac{1}{2}
\mathrm{FT}\left[p^\mathrm{single}(\zeta,\tau)\right]_{\delta/c},
\end{equation}
in agreement with previous theoretical findings \cite{Brok06}. In this
case, the cross-correlation function $g^\times(\tau)$ depends on the
value in $\delta/c$ of the Fourier transform (in $\zeta$) of the
single emitter spectral correlation function
$p^\mathrm{single}(\zeta,\tau)$. Hence, the dynamics of the single
emitter linewidth (as encoded in $p^\mathrm{single}(\zeta,\tau)$) can
be reconstructed through the successive accumulation of
cross-correlation functions $g(\tau)$ at various optical path
differences $\delta$.

Our main point is that the single-emitter component
$g^\mathrm{single}(\tau)$ in the cross-correlation function
$g^\times(\tau)$ does not vanish even when the number of emitters
involved in the experiment becomes very large ($N\gg1$), provided
photoemission occurs with non-Poissonian statistics (i.e.
$g^{(2)}(\tau) \neq1$). Indeed, Equations \ref{Invert1}-\ref{Invert3}
then rewrite as
\begin{equation}\label{TheEq}
g^\times(\tau)=g^{(2)}(\tau)-\frac{1}{2}\mathrm{FT}\left[p^\mathrm{ens}(\zeta)+(g^{(2)}(\tau)-1)p^\mathrm{single}(\zeta,\tau)\right]_{\delta/c},
\end{equation}
where the normalized cross-correlation function $g^\times(\tau)$
contains contributions from both the inhomogeneous ensemble and
single-emitter spectra, with magnitudes of order $1$ and
$g^{(2)}(\tau)-1$ respectively (Figure 3). Information on the single
emitter spectrum will therefore survive ensemble averaging when
emitters, for example, radiate intermittently, as seen on blinking or
transiently-excited emitters - for which $g^{(2)}(\tau\to0)=1+1/n$,
where $n$ is the time-averaged number of emitters significantly
contributing to the ensemble spectrum at any time. The result also
holds for emitters showing complete photon antibunching (i.e.
individually behaving as single-photon sources), for which
$g^{(2)}(\tau\to0)=1-1/n$.

We also note that recording cross-correlation functions
$g^\times(\tau)$ over long durations raises the signal-to-noise ratio,
thus improving the measurement of the homogeneous linewidth - a result
in complete contrast with standard spectroscopy, where longer
accumulation times yield higher signal-to-noise ratios at the expense
of greater inhomogeneous broadening.

\section{Application: unveiling spectral fluctuations in Fluorescence
Correlation Spectroscopy}

Fluorescence Correlation Spectroscopy (FCS) \cite{Webb74} provides a
well-defined framework to illustrate these findings and their
significance. We choose two examples to simulate.  In our first
example, we model an ensemble comprised of single emitters with
differing center frequencies, $\omega_i$, and fixed doublet spectrum
consisting of two delta functions at $\omega_i - \frac{\Omega}{2}$ and
$\omega_i + \frac{\Omega}{2}$. As shown below, the underlying doublet
is easily resolved from the broad ensemble spectrum with our method,
despite the lack of evidence for a doublet in the ensemble emission
spectrum (Figure 4). In our second example, we allow the center
frequency of each doublet to fluctuate in time with a  frequency
$\omega_i(t)$ and demonstrate our ability to observe these spectral
dynamics (Figure 5).

\subsection{Numerical methods}

We modeled FCS experiments by simulating photodetection times and
emission wavelengths for spherical emitters of 2 nm radius freely
diffusing in water at room temperature (diffusion coefficient $D=100$
$\mu$m$^2/$s) and excited by a tightly focused beam forming a
spherical Gaussian excitation spot of width $w_0=200$ nm. The
concentration of the emitters in the solution was adjusted so that a
number of emitters $n=10$ were found in the excitation volume at any
time. Detection of the fluorescence from the excitation spot was set
to a total photodetection rate of $I=10^5$ counts/s.

FCS simulations were performed by generating three-dimensional
Brownian motion trajectories for $N$ emitters diffusing in a finite,
cubic-shaped, open volume simulation box centered on the excitation
spot. To do so, the diffusion trajectories of $N$ particles were first
computed in unbounded free-space, and then subsequently put into the
bounded simulation box by considering the latter as the unit cell of a
three-dimensional tiling with periodic boundary conditions. A
concentration of $n=10$ emitters under the laser spot was reached for
a total number of emitters $N=10^5$ in the simulation box. Once the
single-emitter trajectories were computed, each of the $N$ emitters
was assigned a center frequency $\omega_i$ drawn from the underlying
Gaussian ensemble distribution of width $\Delta$ centered on
$\omega_0$. Open volume conditions were enforced by redrawing the
wavelength of an emitter from this distribution every time it reached
a boundary of the simulation box.

Computations of Brownian trajectories can be highly demanding when
high temporal and spatial resolution are required, as is the case in
FCS. However, the problem simplifies by noting that the position of an
emitter needs only to be determined when it creates a photodetection
event. The positions of each emitter were therefore first computed
assuming uniform excitation over the simulation box, i.e. at times
separated by intervals distributed with Poissonian statistics.
Non-uniform excitation over the simulation box was then taken into
account by filtering these photodetection times with a survival
probability $p=\exp(-\textbf{r}^2/2w_0^2)$ given by the Gaussian
excitation profile at the emitter's location $\textbf{r}$. Emitters
far from the excitation spot at a given time do not radiate, and so do
not contribute to the FCS signal, making the computation of their
trajectory at that time unnecessary. The width of the simulation box
(4 $\mu$m) was therefore kept minimal, yet large enough compared to
the excitation spot size $w_0=200$ nm to keep finite simulation box
effects negligible.

Implemented in C (Anjuta 2.4.1) on a personal computer (1 GHz CPU,
Linux), the above procedure typically required a few hours to produce
long streams of photodetection events consisting of more than $10^7$
photodetection times with their associated detected wavelength for
emitters freely diffusing in a liquid environment under focused laser
excitation. The validity of our approach was confirmed by excellent
agreement between the simulated and theoretical intensity correlation
functions $g^{(2)}(\tau)$ expected from FCS experiments on spherical
shaped emitters in water (Figure 3).

\begin{figure}[t]
\centering\includegraphics{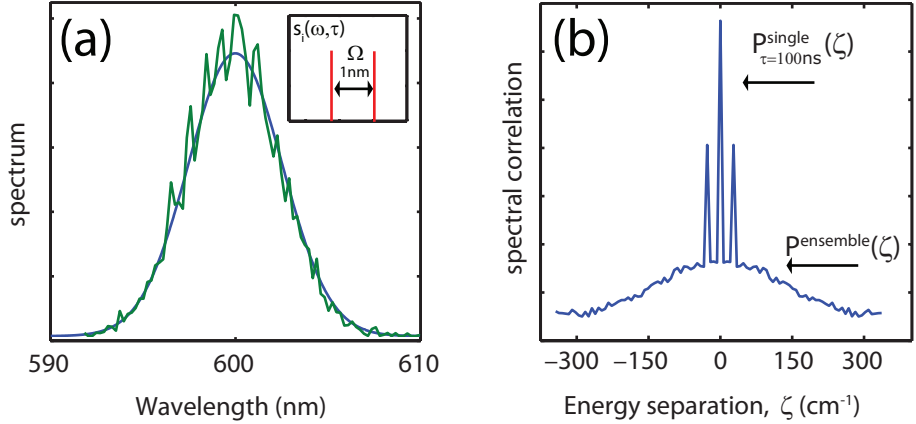}
\caption{Standard spectroscopy versus spectral correlation
measurement. a) Emission spectrum as measured by standard spectroscopy
after an acquisition time of 1 s. The corresponding lineshape
coincides with the average ensemble spectrum, $\langle
S(\omega,t)\rangle$, and shows no evidence of the underlying
single-emitter doublets $s_i(\omega,t)$.  b) Using our method, the
spectral correlation of the underlying doublet,
$p^\mathrm{single}(\zeta,\tau)$, is easily seen on top of a broad
ensemble pedestal, $p^\mathrm{ens}(\zeta)$.  The amplitude of
$p^\mathrm{single}(\zeta,\tau)$ is determined by the intensity
correlation function $g^{(2)}(\tau)$ of the sample emission.}
\end{figure}

\subsection{Line shape of single emitters with static spectra revealed
despite ensemble broadening}

In our first example, ensemble emission was centered at a wavelength
$\lambda_0=600$ nm, with individual transition frequencies $\omega_1,
..., \omega_i, ...$ distributed around $\lambda_0$ with Gaussian
statistics over a FWHM range $\Delta=6$ nm as expected from slight but
significant inhomogeneous broadening. Each single emitter spectrum
consisted of a doublet of monochromatic lines separated by a spectral
width $\Omega=1$ nm.

If measured with a conventional FCS detection setup, the correlation
function $g^{(2)}(\tau)$ of the sample peaks at short timescales
$\tau\ll\tau_D$ (Figure 3), accounting for the fact that the total
detected intensity $I(t)$ fluctuates as emitters continuously enter
and exit the excitation volume with an average diffusion time
$\tau_D=w_0^2/4D=100$ $\mu$s. Here, an emitter therefore diffuses out
of the spot within a duration $\tau_D$ comparable to the average delay
$n/I=100$ $\mu$s between its detected photons, hence contributing to
the total fluorescence signal by a few photons at most. Under such
conditions, the spectrum observed in standard spectroscopy reduces to
its inhomogeneous component - namely a broad Gaussian line of width
$\Delta\gg\Omega$. The doublet in the single emitter spectrum is not
detected (Figure 4a).

We then add the Michelson interferometer in the photodetection path
and scan the interferometer continuously over 10 fringes around
various optical path differences $\delta$. The scanning speed $V$ is
set to 5 fringes/s to access the spectral dynamics at all timescales
$\tau<c/2\omega_0V=200$ ms. Each cross-correlation function
$g^\times(\tau)$ is measured by accumulating photons over 1 minute.

The cross-correlation function $g^\times(\tau)$ now shows a strong
dependence on the optical path difference $\delta$ where it was
recorded, which directly provides the normalized spectral correlation
function $p(\zeta,\tau)=p^\mathrm{ens}(\zeta)+
(g^{(2)}(\tau)-1)p^\mathrm{single}(\zeta,\tau)$ by taking the inverse
Fourier transform in Eq.\ref{TheEq}. Photons separated by durations
$\tau\gg\tau_D$ can not be spectrally correlated, since the population
of emitters in the focal spot undergoes complete renewal over
timescales $\tau\sim\tau_D$, accounting for the fact that
$p(\zeta,\tau)$ then coincides with $p^\mathrm{ens}(\zeta)$. On
timescales shorter than $\tau_D$, emitters generally do not have time
to diffuse out of the excitation spot. Photons separated by durations
$\tau<\tau_D$ have a non-zero probability $g^{(2)}(\tau)-1$ of being
from the same emitter and therefore containing information from a
single emitter. It is at these timescales that the underlying doublet
is revealed, as seen from the triplet of lines of relative amplitude
$\{1/4,1/2,1/4\}$ at frequencies $\{-\Omega,0,+\Omega\}$ produced by
the autocorrelation of a doublet of width $\Omega$ \mbox{(Figure 4b)}.

\begin{figure}[t]
\centering\includegraphics{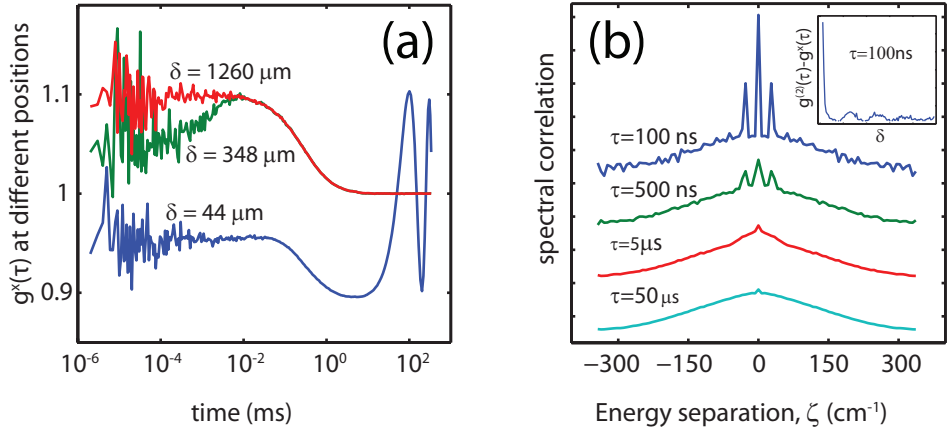}
\caption{Intensity cross-correlations at the ouputs of the scanning
interferometer for a population of emitters undergoing both static and
dynamic spectral broadening. a) $g^\times (\tau)$ as measured at
different interferometer positions $\delta$. The cross-correlation
functions $g^\times (\tau)$ differ from the standard FCS intensity
correlation function $g^{(2)}(\tau)$ at long timescales
$\tau\gg\tau_D$ due to the ensemble spectral correlation
$p^\mathrm{ens}(\zeta)$ and at short timescales $\tau \ll \tau_D$ due
to the time-dependent single-emitter spectral correlation function
$p^\mathrm{single}(\zeta,\tau)$. b) Decay of the $g^\times (\tau)$
with $\delta$ at short timescale $\tau=100$ ns (insert). The
corresponding Fourier transform of $g^{(2)}(\tau)-g^\times (\tau)$ for
increasing values of $\tau$ reveals the time dependent single emitter
spectral correlation, $p^\mathrm{single}(\zeta,\tau)$.}

\end{figure}

\subsection{Line shape of single emitters with a dynamic spectrum}

The method also yields results on single emitters with a dynamic
spectrum. Simulations were performed for emitters with a doublet
spectrum of width $\Omega=1$ nm undergoing both static broadening over
a FWHM range $\Delta=6$ nm and individual dynamic Gaussian spectral
fluctuations with a correlation time $\tau_c=100$ $\mu$s over a broad
FWHM spectral range $\sigma=3$ nm around their center frequencies
$\omega_i$. All other parameters (emitter radius, scanning speed,
detection intensity etc.) were left unchanged from previous section.

In this case, the time-averaged spectrum is a broad Gaussian,
while the time-resolved spectrum is a doublet of separation $\Omega=1$
nm. Here again, standard spectroscopy cannot provide the single
emitter linewidth, as spectral broadening caused by diffusion under
the excitation spot and spectral diffusion of the single emitter
happens at a rate $\tau_D^{-1}+\tau_c^{-1}=20$ ms$^{-1}$ faster than
the average single-emitter photodetection rate $I/n=10$ ms$^{-1}$.

Figure 5a shows $g^\times(\tau)$ for several different interferometer
positions $\delta$. The corresponding patterns in the correlation
functions $g^\times(\tau)$ can be understood as follow. For optical
path differences $\delta$ comparable to the ensemble coherence length
$\Lambda=\Delta/c=40$ $\mu$m, the output intensities are strongly
modulated by the interference pattern of the ensemble spectrum,
correspondingly producing strong intensity anticorrelations between
the interferometer outputs at every timescale $\tau$ ($\delta=44$
$\mu$m, Figure 5a). If we now increase the optical path difference
$\delta$ until the ensemble coherence length $\Lambda$ is exceeded,
fringes emanating from the ensemble spectrum completely vanish
($\delta=348$ $\mu$m, Figure 5a). In this regime, distortions from the
correlation function seen in standard FCS (and in Figure 3)
nonetheless persist due to single-emitter interference phenomena.
Indeed, when plotting $g^\times(\tau)$ as a function of $\delta$, a
beatnote is then evidenced at very short timescales $\tau<100$ ns,
showing that the doublet is resolved over delays
$\tau\ll\min(\tau_c,\tau_D)$, in agreement with our previous
theoretical analysis (Figure 5b, insert). Finally, for very large
optical path differences, all interference phenomena vanish, and the
standard FCS correlation function is actually recovered ($\delta=1260$
$\mu$m, Figure 5a).

Calculating the inverse Fourier transform of
$g^{(2)}(\tau)-g^\times(\tau)$ for different values of $\tau$ clearly
shows the autocorrelation of a temporally evolving doublet - namely a
triplet of intensities $\{1/4,1/2,1/4\}$ in $\{-\Omega,0,+\Omega\}$
superimposed on the broad autocorrelation of the ensemble spectrum
(Figure 5b).

\section{Conclusion}

Our approach overcomes both ensemble and temporal averaging effects in
large populations of single emitters to provide the linewidth of a
single emitter, even if many emitters are detected simultaneously,
with each of them contributing only a few photons to the ensemble
spectrum.

We demonstrated our approach with simulations in conjunction with FCS,
showing the ability to extract a single emitter line shape from an
inhomogeneously broadened ensemble.  We then made that line shape time
dependent and were able to observe the spectral dynamics with high
temporal and spectral resolution.

No assumption was made as to the nature of the excitation beam.
Illustrated here under continuous excitation, our approach applies to
emitters excited by a broadband lamp or a monochromatic laser. Pulsed
excitation is also possible, particularly for the exploration of
spectral correlations (e.g. multi-excitonic spectral lines) occurring
on timescales shorter than the excited state lifetime of the emitters.

\section{Acknowledgments}

This work was supported by the Department of Energy grant number DE-FG02-07ER46454.

\end{document}